\documentclass[showpacs,preprintnumbers,twocolumn,amsmath,amssymb]{revtex4}
\usepackage{graphicx}% Include figure files
\usepackage{dcolumn}% Align table columns on decimal point
\usepackage{bm}% bold math

\begin{document}
%
%\newpage

%\preprint{APS/123-QED}

\title{Repulsion and attraction in high Tc superconductors}
\author{W. LiMing} \email{wliming@scnu.edu.cn}
\author{Haibo Chen, Liangbin Hu}
\affiliation{Dept. of Physics, and Laboratory of Quantum Information Technology, School
of Physics and Telecommunication Engineering, South China Normal University, Guangzhou
510006, China} \keywords{gap symmetry, cuprate, superconductivity}
%\affiliation{$^\dagger$Chinese Science Academy}
\date{\today}
\begin{abstract}
The influence of repulsion and attraction in high-Tc superconductors to the gap functions is studied.  A systematic method is proposed to compute the gap
functions using the irreducible representations of the point group. It is found that a pure s-wave superconductivity
exists only at very low temperatures, and attractive potentials on the near shells
significantly expand the gap functions and increase significantly the critical
temperature of superconductivity.  A strong on-site repulsion
drives the $A_{1g}$ gap into a $B_{1g}$ gap.  It is expected that
superconductivity with the $A_{1g}$ symmetry reaches a high critical temperature due to
the cooperation of the on-site  and  the next-nearest neighbor attractions.
%The gap
%solutions of the gap equation are very stable for a wide range of parameters.
\end{abstract} \pacs{74.20.-z}
 \maketitle

\section{Introduction}

A few decades of study on high-Tc superconductor confirmed the $d_{x^2-y^2}$-wave
symmetry of the gap function at least in cuprate superconductors, relative to the
conventional $s$-wave gap symmetry\cite{Kotlier,Tsuei}. %The observation of the half
%quantum of magnetic flux provides a direct detection to the d-wave gap symmetry in
%cuprates\cite{Tsuei}.
The newly discovered iron-based superconductors have a different gap
symmetry from cuprates, expressed as $\sim \cos k_x \cos k_y$\cite{Yan,Parish}, named as
$s^\pm$-wave. Gap symmetries are believed to be results of point group symmetries of the
superconductor crystals. Tsuei and Kirtley provided a systematic description to the gap
symmetries from the point group theory\cite{Tsuei1}. The relation between the gap
symmetries and the interaction, however, are not clear in literatures. That is to say,
what interaction determines a given gap symmetry is still unknown.

The interaction in high-Tc superconductors is much more complicated than that in
conventional s-wave superconductors. It is widely believed that the on-site repulsion of
strongly correlated electrons drives Cooper pairs from s-wave to d-wave coupling.
Monthous {\it et al}\cite{Monthoux} gave a schematic picture for the interaction in
cuprate superconductors as shown in Fig.1, where an electron locates at the center and another one hopping on
the lattice. Due to the spin fluctuation repulsive
potentials and attractive ones appear alternatively on the lattice sites but decay with
increasing distances. A question is how they affect the gap functions of different symmetries and the
critical temperatures of superconductivity. Understanding this
mechanism may help humans design new superconductors with higher critical temperatures.

\begin{figure}
%\label{tem}
\includegraphics[width=5cm,height=5cm]{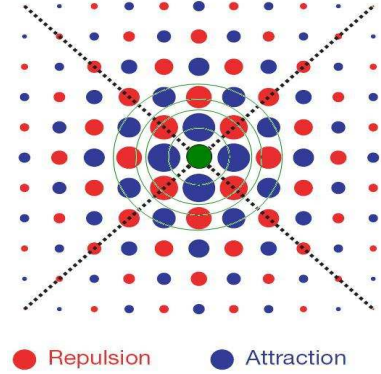}
\caption{Potentials between a quasi-particle at the center of a square lattice (green
spot) and another one moving on the lattice (red or blue spots). The radii of the spots
represent the strength of the potentials\cite{Monthoux}. Each of the four light green circles connects
the neighbors on a shell. From inner to outer the shells are labeled as I, II, II, and IV.}
\end{figure}

In this paper we first reveal the relation between the gap symmetries of a square lattice
and the interactions on the lattice. This will provide a simple method for computing the gap
functions of various symmetries. Using this method we analyze the influence of the
repulsive and attractive potentials on neighboring sites to the gap functions and
critical temperatures of superconductivity. Finally we conclude that superconductivity with the $A_{1g}$ or $B_{1g}$ gap symmetry may have much higher critical temperature than other symmetries, and an on-site attraction favors a $A_{1g}$ gap symmetry but an on-site repulsion favors $B_{1g}$.

\section{Gap function}
The BCS-type pairing Hamiltonian of superconductivity is written as
\begin{align}\label{pair}
H_{pair} &=  {1\over N}\sum_{kk'}V({\bf k'-k})
c^\dagger_{k\uparrow}c^\dagger_{-k\downarrow}c_{-k'\downarrow}c_{k'\uparrow}
\end{align}
where $c^\dagger_{k\sigma},c_{k\sigma}, \sigma=\uparrow,\downarrow$ are the creation and
destruction operators of electrons with spin up and down. Note that a factor $1/N$
has been added in the Hamiltonian. It is necessary for convergence over k-space summation
but missing in literatures for decades. The potential $V({\bf q})$ is expanded into the real space
\begin{align}V({\bf q}) = (1/2)\sum_{\bf m} V({\bf m}) e^{-i{\bf q}\cdot {\bf m}}
\end{align}
where $V({\bf m})$ is the potential energy between an electron at the center and another
one at site ${\bf m}$, as shown in Fig.1.

The central site has the first shell neighbors, i.e., the nearest neighbors($n.n.$), the
second shell neighbors, i.e., the next nearest neighbors($n.n.n.$), the third shell ones,
and the fourth shell ones, etc. In general, only the on-site potential and the potentials
on the four near shells, $V_L, L= 0, 1, 2, 3, 4$, are important. Therefore, these five
parameters determine the pairing Hamiltonian thus the properties of superconductivity.
The gap function is given by the following well-known gap equation\cite{Annett}
\begin{align}
\Delta_{\bf k} &=  -{1\over N}\sum_{{\bf k}'} V({{\bf k}-{\bf k}'}) {\tanh(\xi_{{\bf
k}'}/2kT)\over 2\xi_{{\bf k}'}}\Delta_{{\bf k}'}
\end{align}
where $\xi_{\bf k} = \sqrt{\epsilon_{\bf k}^2 + \Delta_{\bf k}^2}$ is the energy of a
quasi-particle, and $\epsilon_{\bf k}$ is the free-electron energy on a square lattice
given by $\epsilon_{\bf k}=-2t(\cos k_x+\cos k_y) - 4t'\cos k_x\cos k_y-\mu$. Changing
the above gap function from k-space into the real space using a Fourier transformation
$\Delta_{\bf k}=\sum_{\bf m}\Delta({\bf m}) e^{-i{\bf k\cdot m}} $ one obtains
\begin{align}\label{dm}
\Delta({\bf m}) &=  -{V({\bf m})\over 2N}\sum_{\bf m', k'} {\tanh(\xi_{\bf k'}/2kT)\over
2\xi_{\bf k'}}e^{i{\bf k}'\cdot ({\bf m}-{\bf m}')}\Delta({\bf m}')
\end{align}
Compared to the case of k-space the above equation has much smaller dimensions, which is
in general smaller than $21\times21$ as defined by  the 20 neighbors of the four shells
around the central one in Fig.1.  The dimensions can be further reduced up to $5\times 5$
when lattice symmetries are included. In particular, obviously, a unique on-site potential $V_0$
will result in a $k$-independent potential $V({\bf k}) = V_0$ thus the conventional
$s$-wave constant gap function $\Delta(0)$. Dominant neighboring potentials $V_1, V_2,
V_3, V_4$ lead to the unconventional superconductivity, such as the $d$-wave ones.

\section{The symmetries of gap functions on a square lattice}
The gap equation (\ref{dm}) is symmetric about the point group $G$
of the lattice since $\epsilon_{Rk}=\epsilon_k, V({R\bf m}) = V({\bf m}), R\in G$, that
is
\begin{align}\label{drm}
\Delta({R\bf m}) &=  -{V({\bf m})\over 2N}\sum_{\bf m',k'} {\tanh(\xi_{\bf k'}/2kT)\over
2\xi_{\bf k'}}e^{i{\bf k}'\cdot ({\bf m}-{\bf m}')}\Delta({R{\bf m}'})
\end{align}
This shows that the gaps $\Delta({\bf m})$ transform according to the irreducible
representations (IR's) of the point group $G$, i.e.,
\begin{align}
\Delta^{(j)}({R\bf m})=\sum_i T^\alpha_{ij}(R^{-1}) \Delta^{(i)}({\bf m}) \end{align}
where $T^\alpha(R^{-1})$ is the matrix of IR $\alpha$ for group element $R^{-1}$. It
is seen that  gaps $\Delta({\bf m})$ on the same shell differ from each
other at most by a matrix $T^\alpha(R^{-1})$. In particular, they differ only by a phase
factor for one dimensional IR's. This property is represented by the basis functions of
IR's, i.e., gaps $\Delta({\bf m})$ transform under the point group just as the basis
functions do.

%The gap function in a square lattice is the most relevant to the high-temperature
%superconductivity such as cuprates, FeAs series.
A 2D square lattice has a point group symmetry of $D_4$. This group has five IR's $A_{1g},
A_{2g}, B_{1g}, B_{2g}$ and $E{g}$ and their corresponding basis functions as listed in
Table I.
\\ \\
Table I. Irreducible representations and basis functions.
\begin{tabular*}{8cm}{@{\extracolsep{\fill}}llll}
 \hline\hline
 $IR$  &  Basis function& Allowed shells & Wave type\\
\hline
 $A_{1g}$ &$1, (x^2 + y^2)/2$ &0,I-IV& $s + s^\pm$\\
 $A_{2g}$ & $xy(x^2 - y^2)/2$ &IV & $g$\\
 $B_{1g}$ & $(x^2 - y^2)/2$ &I, III, IV& $d_{x^2-y^2}$ \\
 $B_{2g}$ & $xy$ &II,IV & $d_{xy}$\\
 $E{g}$  &$\binom {x+iy} {x-iy}$& I-IV& $p$\\
\hline\hline
\end{tabular*}\\ \\
The first four of these IR's are one dimensional and the last one is two dimensional. The
basis functions of the five IR's listed in Table I vanish except on the
%\begin{figure}
%\label{tem}
%\includegraphics[width=6cm,height=5cm]{lattice.eps}
%\caption{The neighbors of a lattice
%site at the center labeled as 0. The
%curves are guides to the eyes for
%shells of the neighbors.}
%\end{figure}
 allowed shells.  For example, a $g$-wave can only occur on shell IV, and a $d_{xy}$-wave
 on shells II and IV, $etc$.  Then  gaps $\Delta({\bf m})$ are given
 by
\begin{align}\label{dir}
\Delta({\bf m})&= {f_{IR}(x_m, y_m)\over |f_{IR}(x_m, y_m)|} \Delta_{L_{\bf m}}, \quad
{\bf m}\ne 0
\end{align}
where $f_{IR}(x_m, y_m)$ is the value of the basis function at site $(x_m,y_m)$, and
$\Delta_{L_{\bf m}}$ is the magnitude of $\Delta({\bf m})$ on  shell $L_m$.

Using (\ref{dir}) and making an inverse Fourier transformation to $\Delta({\bf m})$ one obtains the gap
functions of every IR,
\begin{align}
\Delta_{A_{1g}}({\bf k})&=\Delta_0 + 2\Delta_1(\cos k_x+\cos k_y)\nonumber\\
&+ 4\Delta_2\cos k_x\cos k_y+2\Delta_3(\cos 2k_x+\cos 2k_y)\nonumber\\
 &+4\Delta_4(\cos 2k_x\cos k_y+\cos k_x\cos 2k_y)\\
 \Delta_{A_{2g}}({\bf k})&= 4\Delta_4(\sin k_x\sin 2k_y-\sin 2k_x \sin k_y)\\
\Delta_{B_{1g}}({\bf k})&=  2\Delta_1(\cos k_x-\cos k_y)\nonumber\\
&+2\Delta_3(\cos 2k_x-\cos 2k_y)\nonumber\\
 &+4\Delta_4(\cos 2k_x\cos k_y-\cos k_x\cos 2k_y)\\
\Delta_{B_{2g}}({\bf k})&=  - 4\Delta_2\sin k_x\sin k_y\nonumber\\
&-4\Delta_4(\sin 2k_x\sin k_y+\sin k_x\sin 2k_y)\end{align}
\begin{align}
 \Delta_{E_{g}}({\bf k}) &= -2 \Delta_1(iu\sin
k_x
-v\sin k_y)\nonumber\\
&-4 \Delta_2(iu\sin k_x \cos k_y -v\cos k_x \sin
k_y)\nonumber\\
 &-4 \Delta_3(iu\sin 2k_x -v\sin
2k_y)\nonumber\\
&-4 \Delta_4[iu(\sin k_x \cos 2k_y+2\sin 2k_x \cos k_y)\nonumber\\
&-v(\cos 2k_x \sin k_y + 2 \cos k_x \sin2 k_y )]
\end{align}
where $\Delta_L, L=0,1,2,3,4$ are the magnitudes of the on-site gap and the gaps on the
four shells, and $u={1\over\sqrt 2}\binom {1} {1}, v={1\over\sqrt 2}\binom {1} {-1}$. A
s-wave gap $\Delta_0$ has been added into  the gap function of $A_{1g}$ since the basis
function of this IR is circularly symmetric. That is to
say, a s-wave gap coexists only with that of the $A_{1g}$ symmetry.  Which gap symmetry is selected by a superconductor is
determined by the lowest condensation energy of the system.

These gap functions are simpler and clearer than the earlier results obtained by  Tsuei
and Kirtley\cite{Tsuei1}. They gave only the first terms of these gap functions.  We will see in the next section
that the first terms may not be  dominant.
It should be emphasized that in general gap functions of
different symmetries do not mix together, unless the symmetry (\ref{drm}) is broken.
Mixing gap functions  were considered due to minor anisotropy in the $a$ and $b$ directions
on $CuO_2$ planes, such as $d_{x^2-y^2} + i d_{xy}$ proposed by Ghosh {\it et
al}\cite{Ghosh} and $s + d_{x^2-y^2}$ by M\"{u}ller {\it et al}\cite{Muller, Maki,
Valls}.

\section{Critical temperatures}

The gap functions of different symmetries are easily worked out through self-consistent
computations to (\ref{dm}) given potential energies $V_L$ on different shells of a square
lattice. It is found that the critical temperatures of superconductivity  of different
gap symmetries are significantly different.

In Fig.2, a pure s-wave gap is created by a pure on-site attractive potential
$V_0=-0.5$ and  gaps with the $A_{1g}$ symmetry are created by the same on-site
potential and other four $ -0.1, -0.2, -0.1, -0.01$ on the four near shells. It is seen
that the s-wave gap exists only below a very low temperature of about $0.00018t$. Small
attractive potentials on the near shells, especially on shell II, extends the on-site gap
a few times and increases the critical temperature more than five times. Therefore,
superconductivity is significantly enhanced by a pairing on the $n.n.n.$ with the
$A_{1g}$ symmetry. This pairing has  two main components $\Delta_0$ and $\Delta_2$. This
leads to a mixing $s+s^\pm$ gap function $\Delta({\bf k}) \approx \Delta_0 + \Delta_2 \cos
k_x \cos k_y$. It has different signs at the center and
$(\pi, 0)$ of the BZ just as the case  in iron pnictides\cite{Graser}.
Further computations show that the potential on shell I,
$V_1$, even if strongly repulsive, e.g. $0.4$, does not influence $\Delta({\bf k})$
significantly. This is similar to the result obtained by Wang and co-workers\cite{Wang} for iron pnictides. This form of gap function is much more robust than the widely assumed form $\cos k_x + \cos k_y$ in literatures\cite{Chubukov}.
\begin{figure}
%\label{tem}
\includegraphics[width=7cm,height=5cm]{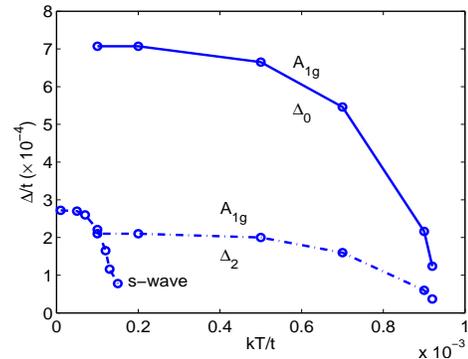}
\caption{Comparison between the gap of a pure s-wave and those of the $A_{1g}$ symmetry.
The pure s-wave gap is created by a pure, on-site, attractive potential $V_0 = -0.5$, and
the gaps of $A_{1g}$ by the five potentials $V_L = -0.5, -0.1, -0.2, -0.1, -0.01 $. Band
parameters are set to be $t = 1.0, t' =-0.125, \mu=-0.527$ for a hole concentration of
0.125.}
\end{figure}

When the attractive potential on shell II, $V_2$, increases to $-0.5$ the gaps
dramatically expand as shown in Fig.3.  Under the cooperation of attractive $V_0$ and $V_2$ the  gaps increase significantly leading to a dramatic increase of $T_c$
(about 52 times compared to the pure s-wave case). If the on-site
attraction is weak the gap is mainly of a $s^\pm$ wave. Thus the superconductivity is
significantly enhanced by the shell II pairing with the $A_{1g}$ symmetry. This provides
a possibility for researchers to increase the $T_c$ of superconductivity by means of
increasing the attraction on shell II atoms. This is just the case of the iron-based
superconductors. It is expected that these superconductors may have much higher critical
temperatures than other series of superconductors.
\begin{figure}
%\label{tem}
\includegraphics[width=6cm,height=5cm]{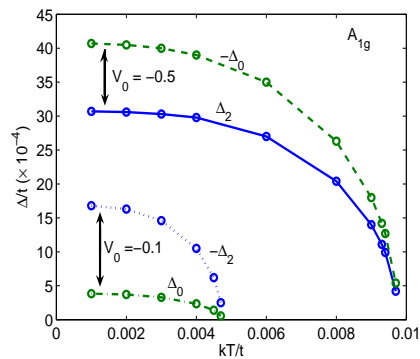}
\caption{Gaps with on-site potentials $-0.1$ and $-0.5$. The four-shell potentials are
$-0.1, -0.5, -0.2, -0.01$. Band parameters are $t = 1.0, t' =-0.125, \mu=-0.527$.}
%= 1.0, t_2 = 0.0, U_0 = 0.0, U_1 = -1., U_2=0.0, \mu=0.0$(left panel) and with the
%$s_{x^2y^2}$ symmetry due to the $n.n.n.$ coupling $t_1 = 1.0, t_2 = 0.0, U_0 = 0.0, U_1
%= 0., U_2=-1.0, \mu=0.0$(right panel). }
\end{figure}
\begin{figure}
\includegraphics[width=6cm,height=5cm]{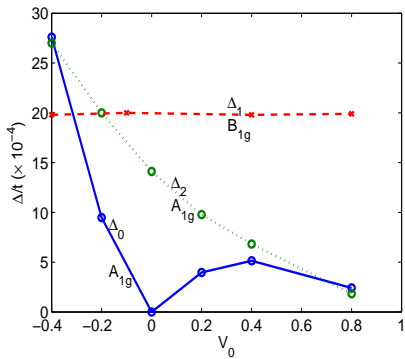}
\caption{Gaps under different on-site potentials $V_0$. Potentials on the four shells are
$-0.3,-0.5, -0.2, -0.01$. $t = 1.0, t' =-0.125, \mu=-0.527$.}
\end{figure}

In cuprates, however, there is a strong on-site Coulombic repulsion due to the strong
correlation effect. Thus a on-site
pairing is   energetically expensive in these superconductors. The gaps with the $A_1$ and $B_1$
symmetries under different on-site potentials are shown in Fig.4.  It is seen that the
on-site repulsion strongly reduces the gaps with the $A_{1g}$ symmetry but holds those
with the $B_{1g}$ symmetry. Therefore, under a strong on-site repulsion a pairing with
the $B_{1g}$ symmetry, i.e. a $d_{x^2-y^2}$ wave, is much more energetically favorable. This is just the case in
cuprates which display a d-wave superconductivity.

\begin{figure}
\includegraphics[width=7cm,height=5cm]{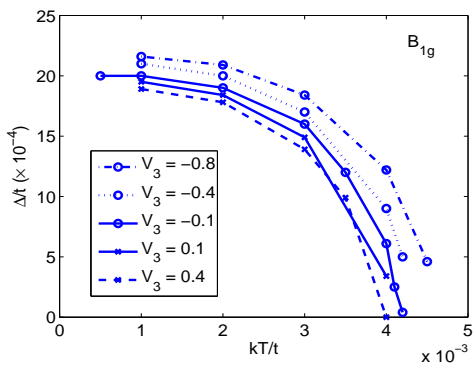}
\caption{Gap $\Delta_1$ with the $B_{1g} $ symmetry vs potential on shell III. Other
potentials are $-0.1, -0.5, -0.2,-0.01$ except $V_3$. $t = 1.0, t' =-0.125, \mu=-0.527$}
\end{figure}

A question is whether this $B_{1g}$ pairing can be further enhanced by increasing
attractive potentials on other shells.  It is found through computations that the
potential on shell II, no matter attractive or repulsive, does not affect the $B_{1g}$
pairing, but  $V_3$ has a weak influence to the gap,  as shown in Fig.5. $\Delta_1$
increases only slightly with more attractive $V_3$ but even a repulsive $V_3$ does not break
the $B_{1g}$ pairing. %Therefore, this pairing is determined almost completely by the
%attractive potential on shell I. This is quite different from the case of the $A_{1g}$
%pairing.

Gaps with other symmetries $A_{2g}$, $B_{2g}$ and $E_g$ are verified through computations
to be small and result in low critical temperatures. Superconductivity with the $A_{1g}$
and $B_{1g}$ symmetries is most expectable to go into the higher temperature region.

\section{Conclusion}
In this work we studied the influence of repulsion and attraction in high temperature superconductors to the gap functions and transition temperatures.
We proposed a systematic method to
compute the gap functions using the irreducible representations of the crystalline point group. Then
we analyze the gap functions  with various symmetries for different
potential energies at different temperatures. It is found that a pure s-wave
superconductivity exists only at very low temperatures, and attractive potentials on the
near shells significantly expand the gap functions and increase dramatically the critical
temperatures. Especially the $A_{1g}$ gap is  expanded by an on-site attractive potential
and that on the next nearest neighbors, thus increases the critical temperature
 for tens of times.  A strong on-site repulsion, however, blows up the $A_{1g}$ gap
but keeps a $B_{1g}$ gap survive.  It is expected that superconductivity
with the $A_{1g}$ symmetry has a high critical temperature.

%\section{acknowledgment}
 This work was supported by the National Natural Science Foundation of
China (Grant No. 10874049),  the State Key Program for Basic Research of China (No.
2007CB925204) and the Natural Science Foundation of Guangdong province ( No. 07005834 ).

\end{document}